\documentclass[utf8]{frontiersSCNS} % for Science, Engineering and Humanities and Social Sciences articles

\setcitestyle{square} % for Physics and Applied Mathematics and Statistics articles
\usepackage{url,hyperref,lineno,microtype,subcaption}
\usepackage[onehalfspacing]{setspace}

\def\keyFont{\fontsize{8}{11}\helveticabold }
\def\firstAuthorLast{Wang {et~al.}} %use et al only if is more than 1 author
\def\Authors{J. Wang\,$^{1,2,*}$, D. W. Xu\,$^{1,2}$ and J. Y. Wei\,$^{1,2}$}
\def\Address{$^{1}$Key Laboratory of Space Astronomy and Technology, National Astronomical Observatories, Chinese Academy of Sciences, Beijing
100012, China \\
$^{2}$School of Astronomy and Space Science, University of Chinese Academy of Sciences, Beijing, China}
\def\corrAuthor{J. Wang}

\def\corrEmail{wj@bao.ac.cn}

\begin{document}
\onecolumn
\firstpage{1}

\title[SDSS\,J0901+6243: A NEW OFeLoBAL]{SDSS\,J090152.05+624342.6: A NEW ``OVERLAPPING-TROUGH'' FeLoBAL QUASAR AT Z$\sim2$} 

\author[\firstAuthorLast ]{\Authors} %This field will be automatically populated
\address{} %This field will be automatically populated
\correspondance{} %This field will be automatically populated

\extraAuth{}% If there are more than 1 corresponding author, comment this line and uncomment the next one.
%\extraAuth{corresponding Author2 \\ Laboratory X2, Institute X2, Department X2, Organization X2, Street X2, City X2 , State XX2 (only USA, Canada and Australia), Zip Code2, X2 Country X2, email2@uni2.edu}

\maketitle

\begin{abstract}

%%% Leave the Abstract empty if your article does not require one, please see the Summary Table for full details.
\section{}
We here report an identification of SDSS\,J090152.04+624342.6 as a new ``overlapping-trough''  iron low-ionization broad absorption line
quasar at redshift of $z\sim2.1$. No strong variation of the broad absorption lines can be revealed 
through the two spectra taken by the Sloan Digital Sky Survey with a time interval of $\sim6$yr.
Further optical and infrared spectroscopic study on this object is suggested. 

\tiny
\keyFont{ \section{Keywords:} quasars: absorption lines --- quasars: individual (SDSS\,J090152.05+624342.6)
--- galaxies: active --- quasars: emission lines ---line: identification}   %All article types: you may provide up to 8 keywords; at least 5 are mandatory.
\end{abstract}

\section{Introduction}

Broad absorption line (BAL) quasars are the objects whose spectra show gas absorptions with a blueshfited outflow 
velocity from $2000\mathrm{km\ s^{-1}}$ up to 0.1$c$ (Weymann et al. 1991).  Although the detailed physics of the outflow is still an open issue (e.g., Fabian 2012),
the outflow is believed to play an important role in the  coevolution of the supermassive blackhole (SMBH) and its host galaxy, which
is firmly established in local AGNs (see Heckman \& Best 2014 for a review) by either expelling circumnuclear gas (e.g., Woo et al. 2017;
Kormendy \& Ho 2012) or 
triggering star  formation through gas compressing (e.g., Zubovas et al. 2013; Ishihashi \& Fabian 2014).

Previous studies, especially the ones based on the Sloan Digital Sky Survey (SDSS, York et al. 2000), indicate that at low and intermediate redshift the fraction of 
BAL quasars is about 20-40\% (e.g., Hewett \& Foltz 2003; Reichard et al. 2003; Trump et al. 2006; Dai et al. 2008;
Urrutia et al. 2009; Knigge et al. 2008; Scaringi et al. 2009), depending on the selection method.  
About 90\% of the BAL quasars are characterized by only high-ionized broad absorptions lines (HiBALs, e.g., CIV, SiIV, NV, OVI).   
The low-ionized absorption lines, such as MgII and AlIII, are identified in the so-called LoBAL quasars with a fraction of $\sim10\%$. 
Among the LoBAL quasars, a small subset ($\sim1\%$ of BAL quasars) of objects are classified as FeLoBAL quasars 
according to their FeII and/or FeIII 
absorption lines (Zhang et al. 2010; Gibson et al. 2009; Hazard et al. 1987; Hall et al. 2002; Yi et al. 2017; Brunner et al. 2003).

Although the physical origin of BAL quasars is originally ascribed to the orientation effect (e.g., Weymann et al. 1991; Goodrich \& Miller 1995;
Gallagher et al. 2007),  the higher reddening in BAL quasars than in non-BAL quasars motivate a lot of studies 
to try to understand if
BAL quasars are 
young AGNs, in which the FeLoBAL quasars with the highest reddening and column density are possible transitional quasars from a dust-obscured 
AGN to a unobscured one. Mudd et al. (2017) recently identified the first post-starburst FeLoBAL quasar DES\,QSO\,J0330-28 at a redshift of 0.65.

In this paper, we report an identification of SDSS\,J090152.04+624342.6 as a new unusual FeLoBAL quasar  with 
``overlapping-trough'' (OFeLoBAL quasars) at $z\sim2.1$.

\section{Spectroscopic Identification}

\subsection{History of SDSS\,J090152.04+624342.6}

SDSS\,J090152.04+624342.6 was serendipitously extracted from the Sloan Digital Sky Survey (SDSS, York et al. 2000) Data Release 7 spectroscopic catalog,
when we examined the spectrum of the ``unknown'' objects one by one by eye.  
%In the SDSS Data Release 7 spectroscopic catalog,  the object was first classified as a ``unknown'' object. 
The object was then classified
as a quasar at $z=2.09$ in the $7^{\mathrm{th}}$ SDSS Quasar Catalog (Schneider et al. 2010; Shen et al. 2011) by
identifying the broad emission line at the red end as MgII$\lambda2800$.
With a new spectroscopic observation, the redshift was recently (and improperly) updated to $z=6.389420\pm0.000594$ by the pipelines of
SDSS Data Release 13\footnote{http://www.sdss.org/dr13/data\_access/bulk/}
through an identification of the peak as Ly$\alpha$ emission line. Figure 1 shows the observer-frame spectrum of SDSS DR13 and 
that of DR7. In fact, by assuming a redshift of $z\sim6$, 
the object shows abnormally significant emission blueward
of the Lyman limit at a rest-frame wavelength of $\sim6500\AA$ (see the typical spectra of the 
high-redshift quasars at $z\sim6$ in Fan et al. 2006 and Wu et al. 2015 and references therein).

\begin{figure}[ht!]
\begin{center}
\includegraphics[width=0.8\textwidth]{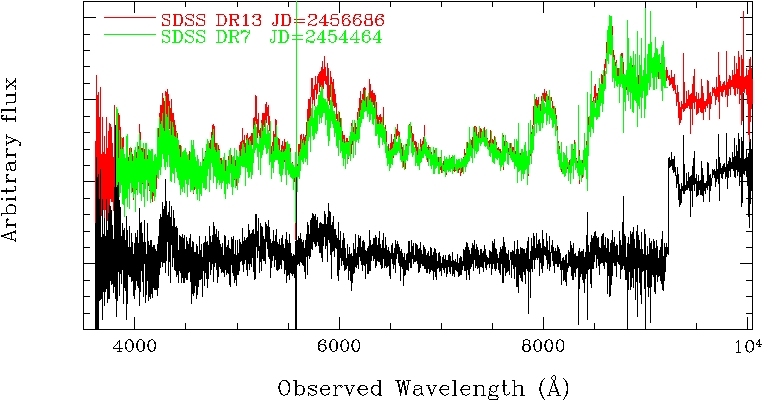}% This is a *.jpg file
\end{center}
\caption{The spectra taken from SDSS DR13 and that from SDSS DR7. 
Both spectra are shown in observer frame. The bottom black curve shows the differential spectrum that is 
vertically shifted by an arbitrary amount for visibility.
}\label{fig:1}
\end{figure}

\subsection{Data Reduction}

The spectral analysis is performed as follows by the IRAF packages\footnote{IRAF is distributed by the National Optical Astronomy Observatory, which
is operated by the Association of Universities for Research in Astronomy, Inc.,
under cooperative agreement with the National Science Foundation.}.
The 1-Dimensional spectra of the object taken from SDSS DR13 is corrected for the Galactic extinction
basing upon the V-band extinction taken from  Schlafly \& Finkbeiner (2011).
An $R_V=3.1$ extinction law (Cardelli et al. 1989) of the MilkyWay is adopted in the correction.

\subsection{Identification of a New OFeLoBAL Quasar}

Both spectra of the object taken from SDSS show an abrupt drops in flux at around the observer frame wavelength of $\lambda\sim8000\AA$
and many ``features'' blueward of the drop, which  
closely resemble the spectra of the unusual OFeLoBAL quasars discovered in previous studies, such as SDSS\,J0300+0048 ($z=0.89$),
SDSS\,J1154+0300 ($z=1.458$), Mark 231, FIRST\,1556+3517 and FBQS\,1408+3054 (e.g., Smith et al. 1995; Becker et al. 1997,2000; 
White et al. 2000; Hall et al. 2002). In the OFeLoBAL quasars, the abrupt drops are caused by 
a blueshifted absorptions due to MgII$\lambda\lambda2796,2803$
and MgI$\lambda2852$, and almost no continuum windows can be identified blueward of the MgII emission because of the overlapping 
troughs mainly due to the FeII and FeIII absorptions. 

Figure 2 shows the rest-frame spectrum of the object, along with our identification of both emission and absorption features. 
By ascribing the peak at the red end of spectrum as an emission from the MgII$\lambda\lambda2796,2803$ doublets,
the systematic redshift of the object is inferred to be $z=2.09$ which is 
consistent with the previous claims in SDSS DR7 quasar catalog (e.g., Schneider et al. 2010; Shen et al. 2011; Wu et al. 2012). 
In fact, this redshift allows us to accurately predict the wavelength of not only the broad emission redward of the MgII emission,
but also the CIV$\lambda1549$ and possible NeV$\lambda1240$ emission features, although the CIII]$\lambda1909$ emission commonly 
appearing in the
quasar's spectra is hard to be identified in this object. The two bumps redward of the MgII emission are identified to be a blend of 
the HeI$\lambda2945$+FeII$\lambda2950$ (UV60 and UV78) complex and a blend of the optical FeII complex at around 3200\AA\ (i.e., Opt7 and Opt6).  

The spectrum blueward of the MgII emission is dominated by multiple overlapping troughs with a redshift of $\sim1.98$.
Again, the redshift accurately predicts the wavelength of the absorptions blueward of the MgII emission. 
The onset of the troughs is a strong MgI$\lambda2857$ absorption followed by damped MgII$\lambda\lambda2796,2803$ absorptions. 
An evident residual flux at high-velocity end of the MgII trough enables us to argue a presence of FeII$\lambda2750$ (UV62 and UV63) 
absorptions, which is followed by the absorption features of FeII UV1 and UV2.  
With the redshift of $\sim1.98$, the troughs at middle of the spectrum are identified as the absorptions due to MgI+ZnII+CrII+FeII UV48, AlIII$\lambda\lambda$1854,1862 
and AlII$\lambda1671$, which are all common in the spectra of FeLoBAL quasars. Finally, two troughs due to SiII$\lambda1527$ (UV2) and 
SiIV$\lambda\lambda$1394,1402 absorptions can be identified at the predicted wavelengths at the blue end of the spectrum. 

\begin{figure}[ht!]
\begin{center}
\includegraphics[width=0.8\textwidth]{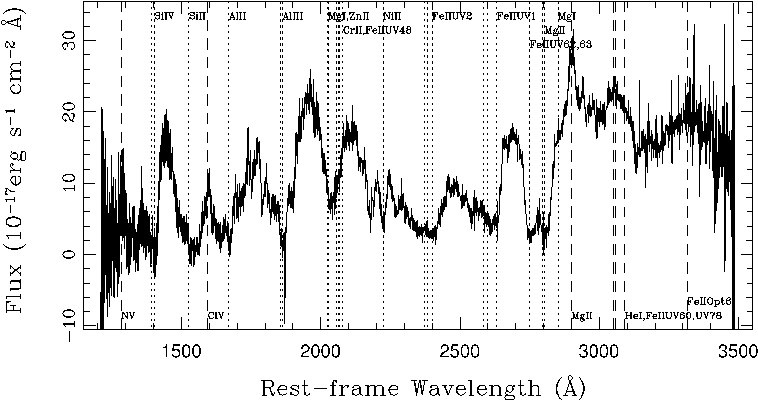}% This is a *.jpg file
\end{center}
\caption{The SDSS DR13 rest-frame spectrum of the object based on the redshift of the absorption features of $z=1.98$. The long 
dashed lines from top to bottom marks the predicted wavelengths of the identified emission features, and the short dashed lines
the wavelengths of the absorptions. The emission and absorption features are labeled at bottom and top of the figure, respectively.   
}\label{fig:2}
\end{figure}

\section{Non-variation of the New OFeLoBAL Quasar}

Significant variation of BALs, including a complete disappearance, with a time scale of 1-10 yr in the quasar rest-frame have been reported in 
the previous studies (e.g., Hall et al. 2011; Filiz Ak et al. 2012. 2013; Joshi et al. 2014; Zhang et al. 2011, 2015; Vivek et al. 2012, 2014).
The significant variation can be explained by a variation of either the ionizing power (e.g., Trevese et al. 2013)
or the covering factor due to a cloud transiting the line-of-sight (e.g., Hall et al. 2011). By comparing the variability of 
OFeLoBAL and non-OFeLoBAL quasars, Zhang et al. (2015) claimed a prevalence of strong BAL variation in the OFeLoBAL quasars rather than in
the non-OFeLoBAL ones, which allows the authors to argue that the troughs in OFeLoBAL quasars are resulted from dense outflow gas closer to the 
central SMBH. 

SDSS\,J090152.04+624342.6 has been observed twice by SDSS with a time interval of $\sim6$ yr, which corresponds to a rest-frame 
time of $\sim2$ yr. The two spectra are compared in Figure 1, 
along with a difference spectrum. The difference spectrum is obtained by a direct subtraction of the 
two spectra at the different epochs, since they are matched very well redward of the MgII line emission. 
One can see from the figure that no significant variation can be identified in the object through
a comparison of the two SDSS spectroscopic observations. The invariant of the spectra of the object suggests a rest-frame life time of 
its BAL structure being no shorter than 2yr. The knife-edge model in Capellupo et al. (2013) gives a simple relation of the
crossing velocity $\upsilon$ of the absorber of
$\upsilon_{\mathrm{cross}}=\Delta A D/\Delta t$, where $\Delta A$ is the fraction of the continuum region crossed by the
absorber, and $D$ the diameter of the continuum region. With the typical values of $\Delta A=0.1$ and $D=10^{-3}$pc (e.g., Capellupo et al. 2013;
McGraw et al. 2015), the 
invariant of the BAL structure of the object within a rest-frame time of 2yr suggests a 
crossing velocity $\upsilon_{\mathrm{cross}}<5\times10^3\mathrm{km\ s^{-1}}$.

\section{Conclusion and Future Study}

SDSS\,J090152.04+624342.6 is identified as a new OFeLoBAL quasar at $z\sim2.1$. The spectra taken by SDSS at two epochs with a time interval of 
6 yr do not show significant variation of its BAL. Further infrared spectroscopic observation is necessary for confirming the redshift determination, studying the host galaxy stellar 
population and estimating BH viral mass through Balmer emission lines. Based on the redshift of $z\sim2.1$, the H$\beta$ line which is 
traditionally used for BH mass estimation, is redshifted to 1.5$\mu$m  at observer frame. And also, further optical spectroscopic and 
photometric monitor is useful for revealing significant BAL variation in the object.

\section*{Conflict of Interest Statement}
%All financial, commercial or other relationships that might be perceived by the academic community as representing a potential conflict of interest must be disclosed. If no such relationship exists, authors will be asked to confirm the following statement: 

The authors declare that the research was conducted in the absence of any commercial or financial relationships that could be construed as a 
potential conflict of interest.
\section*{Author Contributions}

JW initiated the study, conducted data reductions, and wrote the manuscript. DWX and JYW contributed to the discussions and
manuscript preparation.

\section*{Funding}

This study is supported by the National Natural Science
Foundation of China under grants 11473036 and 11773036, and by the National Basic Research Program of 
China (grant 2009CB824800).

\section*{Acknowledgments}

The authors would like to thank the anonymous referee for
very useful comments and suggestions for improving the
manuscript. This study uses the SDSS archive data that was
created and distributed by the Alfred P. Sloan Foundation.
The author would like to thank the anonymous referee from journal of ApJ letter
who pointed out our initial mistake and gave us very useful suggestions.

%\section*{Supplemental Data}
% \href{http://home.frontiersin.org/about/author-guidelines#SupplementaryMaterial}{Supplementary Material} should be uploaded separately on submission, if there are Supplementary Figures, please include the caption in the same file as the figure. LaTeX Supplementary Material templates can be found in the Frontiers LaTeX folder 

\bibliographystyle{frontiersinSCNS_ENG_HUMS} % for Science, Engineering and Humanities and Social Sciences articles, for Humanities and Social Sciences articles please include page numbers in the in-text citations

\end{document}